\documentclass[useAMS,usenatbib]{mn2e}
\usepackage{graphicx,amssymb,txfonts,pstricks,color,xspace}

\title[....]{X-rays as dominant excitation mechanism of  [Fe\,\textsc{ii}] and $\rm H_{2}$ emission lines in active galaxies} 
\author[Dors  et al.]
{Oli L. Dors Jr.$^{1}$\thanks{E-mail:olidors@univap.br}, Rogemar
  A. Riffel$^{2}$,  M\'onica V.\ Cardaci$^{3,4,5}$, Guillermo F. H\"agele$^{3,4,5}$, 
  \newauthor \^Angela C. Krabbe$^1$, Enrique P\'erez-Montero$^{6}$,  Irapuan Rodrigues$^{1}$\\
$^1$ Universidade do Vale do Para\'iba, Av. Shishima Hifumi, 2911, Cep
12244-000, S\~ao Jos\'e dos Campos, SP, Brazil\\
 $^2$ Universidade  Federal de Santa Maria, Av.  Roraima, 1000, Cep
97105-900, Santa Maria, Brazil\\
$^3$ Consejo Nacional de Investigaciones Cient\'ificas y T\'ecnicas (CONICET), Argentina.\\
$^4$Facultad de Ciencias Astron\'omicas y Geof\'{\i}sicas, Universidad Nacional de la La Plata, Paseo del Bosque s/n, 1900 La Plata, Argentina.\\
$^5$ Departamento de F\'{\i}sica Te\'orica, C-XI, Universidad Aut\'onoma de Madrid, 28049 Madrid, Spain.\\
$^6$ Instituto de Astrof\'{\i}sica de Andaluc\'{\i}a (CSIC), PO Box 3004, 18080 Granada, Spain}

\begin{document}


\pagerange{\pageref{firstpage}--\pageref{lastpage}} 

\maketitle

\label{firstpage}

\begin{abstract}

We  investigate the excitation mechanisms of near-infrared [Fe\,{\sc ii}] and $\rm H_{2}$ 
emission lines observed in Active Galactic Nuclei (AGNs). We built a
photoionization model grid   considering a two-component continuum, one 
accounts for the Big Bump component peaking at $\rm 1 Ryd$  
and another represents the X-ray source that dominates   the continuum
emission at high energies. Photoionization models  
considering as ionizing source a spectral energy distribution   obtained from
photometric data  of the Sy 2 Mrk\,1066 taken from the  literature were considered.
Results of these models were compared with     a large sample of observational long-slit and
Integral field Unit (IFU) spectroscopy data of the nuclear region 
for a sample of active objects. 
We found that the correlation between the
observational [Fe\,{\sc ii}]$\lambda$1.2570$\,\mu$m/Pa$\beta$ vs.  H$_2\lambda$2.1218$\,\mu$m/Br$\gamma$  
  is well reproduced by our models as well as the relationships that
involve the H$_2$ emission line ratios observed in the spectroscopic data.
We conclude that the heating by X-rays produced by active nuclei
can be considered  a common and very important mechanism of excitation of [Fe\,{\sc ii}] and $\rm H_{2}$.
\end{abstract}


\section{Introduction}

The excitation of the Narrow Line Region  of Seyfert (Sy) galaxies can reveal how radiation and mass
outflows from the nucleus interact with circumnuclear gas.  In particular, near-infrared (hereafter near-IR) 
observations are a powerful tool to investigate this issue, because the obscuration --which can affect the optical
morphology of the emitting gas region-- is less important at these wavelengths \citep{mulchaey96,ferruit00}. 
Relevant emission-lines in the near-IR include [Fe\,{\sc ii}]\,$\lambda\,1.2570\,\mu$m and
$\lambda\,1.6440\,\mu$m, H{\,\sc i} lines such as Pa$\beta$, and  H$_2$ at $\lambda\,1.9576\,\mu$m, $\lambda\,2.1218\,\mu$m, and $\lambda\,2.3085\,\mu$m, 
which can be used to map the gas kinematics and excitation \citep[e.g.][]{riffel11b,m1066}. Nevertheless, the dominant excitation mechanisms of the  [Fe\,{\sc ii}] and H$_2$ 
emission lines in the central regions of  active galaxies are  still unclear and have been the subject of several recent studies 
\citep[e.g.][]{m1066,riffel08,eso428,n4151,hicks09,sanchez09,ramos-almeida09,rodriguez-ardila05,rodriguez-ardila04,davies07}. 

The H$_2$ can be excited by two mechanisms: (i) fluorescent
excitation through absorption of soft-UV photons (912--1108\,\AA) in
the Lyman and Werner bands, existing both in star-forming regions and
   surrounding  the Active Galactic Nuclei (AGNs)
\citep{black87} and (ii) collisional excitation  due to the heating of the gas by shocks, the interaction of a radio jet with the interstellar medium \citep{hollenbach89}, 
or by X-ray photons from the central AGN \citep{maloney96}.  Several studies 
based on intensity-line ratios \citep{m1066,n4151,rodriguez-ardila05,rodriguez-ardila04} have shown 
that  collisional excitation processes dominate the H$_2$ emission surrounding AGNs. However,
  which is the dominant 
process is an open question. \citet{veilleux97}, using J and K-band spectra of a sample of 33 Sy\,2 galaxies, 
found that shocks associated with nuclear outflows are a likely source of both
[Fe\,{\sc ii}] and $\rm H_{2}$ emission
rather than circumnuclear starbursts, as suggested by \citet{quillen}.

  For the [Fe\,{\sc ii}] emission, the [Fe\,{\sc ii}]\,$\lambda\,1.2570\,\mu$m/Pa$\beta$ line ratio
is  generally used to investigate the main mechanism of excitation. The
value   of this   line ratio
is controlled by the quotient of
the volumes of partially   and fully ionized gas regions, with
[Fe\,{\sc ii}] emission being excited in the partially ionized gas \citep{mouri90,mouri93,rodriguez-ardila05,m1066,riffel08,eso428,n4151}. 
Such zones in AGNs are created by X-ray emission \citep[e.g.][]{simpson96} and/or shock heating of the gas
by mass outflows from the nuclei which interact with the ambient clouds \citep[e.g.][]{forbes93}. 
This problem was  addressed by  \citet{mouri00}, 
who compared the values of the line ratios
[Fe\,{\sc ii}]\,$\lambda\,1.2570\,\mu$m/Pa$\beta$ and [O\,{\sc i}]\,$\lambda\,6300$\,\AA/H$\beta$ predicted by models, considering 
photoionization and shock heating, with those observed in a sample of AGNs and Starburst galaxies. These authors 
  pointed out that in AGNs, X-ray heating is the   most important
[Fe\,{\sc ii}] excitation mechanism.  
However, \citet{rodriguez-ardila04},
using near-IR spectroscopy of a sample of galaxies obtained with the Infrared Telescope Facility,
found that X-ray excitation is enough to explain the H$_2$ emission and part of the  [Fe\,{\sc ii}] emission 
observed in Sy\,1 galaxies, but fails to explain  the emission of these elements in
Sy\,2. For these objects, a combination of shocks   and circumnuclear star-formation
 is required to explain these emissions.  Moreover, it is not
clear   whether the [Fe\,{\sc ii}]
and  H$_2$ are excited by the same mechanism. \citet{rodriguez-ardila04} found a correlation between 
the [Fe\,{\sc ii}]\,$\lambda\,1.2570\,\mu$m/Pa$\beta$ and the
H$_2$\,$\lambda\,2.1218\,\mu$m/Br$\gamma$ ratios,   indicating that
  both sets of lines may be originated by  a single dominant mechanism.
However,   high spatial resolution  spectroscopy data from 
Integral Field   Unit (IFU)  of active galaxies indicate that the H$_2$ 
and the [Fe\,{\sc ii}] emitting gas have distinct flux distributions and
kinematics, with the former being considered a tracer of the  
feeding of the AGN and the latter a tracer of its feedback
\citep{m1066,riffel09,riffel08,n4151,hicks09,sanchez09}. This result
indicates that the lines of these elements are formed 
in distinct regions.  
Although  several works have investigated the 
excitation origins of H$_2$ and the
[Fe\,{\sc ii}], it is still unknown   whether a common
mechanism can  excite  these elements. 
    Fortunately, a large number of near-IR data of AGNs are currently
available in the literature, which enables an extensive comparison with models 
yielding a more reliable conclusion about the likely dominant excitation mechanism
of these emission lines. 

In this paper, we combined near-IR data of Sy galaxies obtained with 
IFU and long-slit spectroscopy with photoionization models   to
investigate the   origin of the H$_2$ and [Fe\,{\sc ii}]. In Section
~\ref{obs}, we describe the observational data used in the analysis. The
modelling procedures are presented in  Sect.~\ref{phot}. In Sect.~\ref{diag},
the diagnostic diagrams used to compare the observational data with our
 model predictions are described. 
Results and discussion are presented in Sects.~\ref{res} and
\ref{disc}, respectively. A conclusion of the outcome is
given in Sect.~\ref{conc}.

\section{Observational data}
\label{obs}

   We compiled from the literature observational data of the nuclear region of
active galaxies in the near-IR and optical spectral range obtained with
long-slit and IFU spectroscopy. The selection criterion was the presence of
bright infrared emission lines in their spectra. These data are described below.

\subsection{Long-slit data}
\label{int}

Near-IR emission line intensity ratios of 35 active galaxies were obtained 
from \cite{rodriguez-ardila04}, \cite{reunanen02}, \cite{knop01}, and
\cite{rogerio06}. This sample  comprises long-slit data of 13 Sy\,1 and 21
Sy\,2 galaxies, along with 1 Quasar. The intensities of the near-IR [Fe\,{\sc ii}] and H$_2$
emission lines observed in these objects were compared with our
photoionization models.
We also used the [O\,{\sc iii}]\,$\lambda$\,5007\,\AA/H$\beta$ and [O\,{\sc
i}]\,$\lambda$\,6300\,\AA/H$\alpha$ line intensity ratios of about 600\,000
emission-line galaxies listed in the MPA/JHU Data catalogue of the Sloan
Digital Sky Survey DR7 release (available 
at http://www.mpa-garching.mpg.de/SDSS/DR7/).

\subsection{IFU data}

  For this study we selected two Sy\,1 galaxies, Mrk\,1157 and NGC\,4151,
and two Sy\,2 galaxies  (ESO\,428-G14 and Mrk\,1066). All of them were previously
observed by our group using the IFU spectrographs of the Gemini telescopes.
We selected these objects  because  both J- and K-band spectroscopic data are
available.
The  observations of Mrk\,1066, Mrk\,1157, and NGC\,4151 were   performed
using the Near-IR Integral field Spectrograph \cite[NIFS;][]{mcgregor03} on
Gemini North,
while ESO\,428-G14 was observed with the Gemini Near Infra-Red Spectrograph
\citep[GNIRS;][]{elias98} on Gemini South.

\section{Photoionization model}
\label{phot}

To analyse the [Fe\,{\sc ii}] and $\rm H_{2}$ excitation   mechanisms,
we built a grid of models using the photoionization code Cloudy/08
\citep{ferland98}, and then we compared the line intensity ratios predicted
by them with  those observed.
The spectral energy distribution (SED) of the ionizing source used as input
for the Cloudy code was a two-component continuum 
ranging from $\sim 10^{15}$\,Hz to $\sim 10^{21}$\,Hz. The shape of this SED
is similar to the one observed in typical AGNs for that range.
The first is the Big Bump component peaking at $\rm 1 \: Ryd$ 
with a high-energy and an infrared exponential cutoff and the second one
represents the X-ray source that dominates at high energies and  is
characterized by a power law with an index $\alpha_{x}=-1$. Its
normalization was computed to produce the required value of the 
optical to X-ray spectral index $\alpha_{ox}$. This index 
describes the continuum between 2 keV and 2500\,\AA \, \citep{zamorani}. 
 We assumed the default value of the Cloudy code $\alpha_{ox}=-1.4$, because that is about the
average of the observed values, which are between -1.0 and -2.0, for the entire range of 
observed luminosities of AGNs \citep{miller11, zamorani}. 

  The cosmic ray emission was considered in the models as a second
ionizing source. Cosmic rays heat the ionized gas and   produce
secondary 
ionizations in the neutral gas, which   mostly increase the intensities
of the $\rm H_{2}$ emission lines. We assumed a value of the
$\rm H_{2}$ ionization rate of $10^{-15} \: \rm s^{-1}$, which is 
about the same   rate  found by \citet{mccall03} for a galactic line of
sight.   It is worth  noting that the value of the cosmic ray rate must be
estimated object by object. For example, \citet{suchkov93} found for M\,82 a
cosmic ray rate several times larger than the one in the Milk Way. 
Gamma ray observations of the starburst NGC\,253 by \citet{acero09}
indicate a cosmic ray rate three orders of magnitude larger than   that for the Milky Way. 
Also, molecular data of star forming galaxies, such as Arp\,220, show evidence
for extremely high cosmic ray rates yielded   by the UV emission from supernova remnants 
\citep{meijerink11}.

We computed a sequence of models assuming an electron density $N_{\rm e}= 10^{4} \rm \: cm^{-3}$,
ionization parameter  $U$ in the range  $-4.0 \leq \log U \leq -1.0$ 
defined as $U=Q_{ion}/4\pi R^{2}_{\rm S} n  c$, where $Q_{ion}$  
is the number of hydrogen ionizing photons emitted per second
by the ionizing source, $R_{\rm S}$  is the Str\"omgren radius (in cm), $n$ is the  particle
density (in $\rm cm^{-3}$), and $c$ is the speed of light.
The chosen range of these values for $U$ is typical of narrow-line regions
of Sy galaxies (e.g. \citealt{ferland83}).  The
H$_2$ emission lines are very dependent on the electron density value assumed in the models.
For example,  when  $N_{\rm e}$ varies from $10^{3}$   to  $10^{5} \: \rm cm^{-3}$,
the  logarithm of the H$_2\,\lambda$\,2.1218$\,\mu$m/Br$\gamma$ emission line
intensity ratio span about 2.6\,dex.
The value $N_{\rm e}= 10^{4} \rm cm^{-3}$, assumed in our models, is a mean
value from those considered by \citet{mouri00}. 
We considered in our models three values of 12+log(O/H)= 8.38, 8.69, and 
9.00, which correspond to values of the metallicity 0.5, 1, and 2 times
the solar value published by \citet{allende-prieto01}. 
The abundances of other metals in the nebula were scaled linearly
to the solar metal composition through the comparison of the
oxygen abundances, with the exception of the N and Fe abundances.
The nitrogen abundance   was taken from the relation log(N/O)=log(0.034+120O/H) of Vila-
Costas \& Edmunds (1993).  The Fe/O abundance ratio has a large scatter for a fixed
O/H value \citep{izotov06} and its value is uncertain  because
the Fe and O abundances in grains are poorly known \citep{peimbert10}. Thus,  
we varied   the Fe/O abundance ratio by about 0.7\,dex on each  metallicity.

\begin{table}
\centering
\caption{Fe/O  and O/H gas phase abundances  assumed in the models.}
\label{table2}
\vspace{0.2cm}
\begin{tabular}{@{}c c c c cc @{}}
\hline
      Metallicity (Z/Z$_{\odot}$)     &          12+log(O/H)&     & \multicolumn{3}{c}{log(Fe/O)}   \\	
 \cline{4-6} 
 \cline{1-2}
2                &      9.0        &  &-1.47(a1)       &  -1.94(a2)        &  -2.24(a3)              \\
1                &       8.69      &  & -1.77(b1)      &  -2.24(b2)       &  -2.77(b3)   \\
0.5              &        8.38     &   &-2.15(c1)      & -2.54(c2)      & -2.76(c3)  \\
\hline
\end{tabular}
\end{table}

The presence of internal dust was considered and the grain
abundances (van Hoof et al. 2001) were also linearly scaled with
the oxygen abundance. To take   into account the depletion of refractory elements
onto dust grains, the abundances of Mg,
Al, Ca,  Ni, and Na were reduced by a factor of 10, and Si by a
factor of 2    relative to adopted abundances in each model
in accordance   with  \citet{garnett95}.
In Table~\ref{table2}, 
the O/H and Fe/O abundance values of the gas phase  assumed in the models are shown.
The model of the $\rm H_{2}$ molecule described by \citet{shaw05}
and the  model of the $\rm Fe^{+}$ ion described by \citet{verner99}, which consider
371   energy levels, were assumed in our computations. The outer radius
  of the modelled nebula is that where the temperature falls below 1000\,K.

\section{Diagnostic Diagrams}
\label{diag}

We used four diagnostic diagrams containing   predicted and observed emission line ratios
of the [Fe\,{\sc ii}] , $\rm H_{2}$, [O\,{\sc iii}], and  [O\,{\sc i}]  which are described below.

\begin{figure}
 \centering
 \includegraphics[angle=-90,width=8cm]{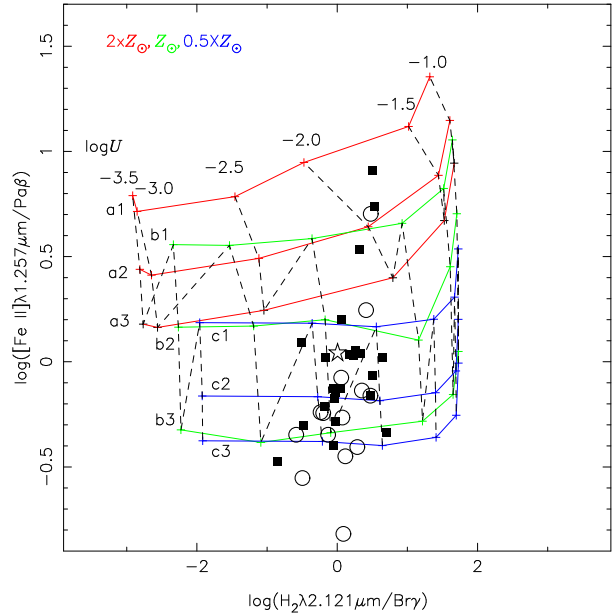}
 \caption{Diagnostic  diagram showing the observational data taken from
   the literature (see Sect. \ref{obs}) and results   from the grid of
   photoionization models (see Sect.\ref{phot}). Solid lines connect curves of
   iso-$Z$, while dotted lines connect curves of iso-$U$. The values of $\log
   U$ and $Z$ are indicated.  The three different lines for each $Z$ 
   correspond to the different assumed values of the Fe/O as indicated by the labels (see
   Table\,\ref{table2}).  Circles, squares, and star represent Sy1, Sy2, 
   and quasar data, respectively. 
   The typical error bar (not shown) of the emission line ratios is about
   10\,\%. }
 \label{fig1}
 \end{figure}

\begin{figure}
 \centering
 \includegraphics[angle=-90,width=8cm]{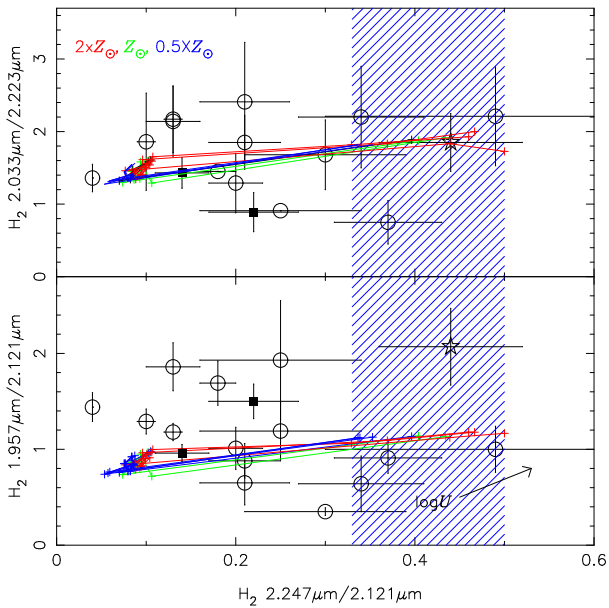}
 \caption{As in Fig.\ \ref{fig1} for $\rm H_{2}$ emission lines. The
   arrow indicates the direction  in which the ionization parameter
   increases.  Circles, squares, and star represent Sy1, Sy2, 
   and quasar data, respectively. The hatched area represents the region occupied by  
   shock model results from \citet{hollenbach89}. }
 \label{fig2}
 \end{figure}

\begin{itemize}

\item $ $ [Fe\,{\sc ii}]\,$\lambda$\,1.2570$\,\mu$m/Pa$\beta$
  vs.\ H$_2\,\lambda$\,2.1218$\,\mu$m/Br$\gamma$   (Fig.\ \ref{fig1}) --- 
Diagnostic diagram suggested by \cite{larkin98} and   \citet{rodriguez-ardila04} to separate galaxies according 
to their level of nuclear activity. Recently, \citet{m1066} constructed  this diagram
with  spatially resolved IFU data of an AGN. Typical values for the nucleus of Sy
galaxies are $0.6\,\lesssim$\,[Fe\,{\sc ii}]\,$\lambda$\,1.2570$\,\mu$m/Pa$\beta$\,$\lesssim\,2.0$ and
$0.6\,\lesssim$\,H$_2\,\lambda$\,2.1218$\,\mu$m/Br$\gamma$\,$\lesssim\,2.0$
\citep{rodriguez-ardila05}.   This [Fe\,{\sc ii}]/Pa$\beta$ is very dependent on
the Fe/O abundance while the H$_2$ emission lines are dependent on the ionization parameter.
 
\item $ $ H$_2\lambda$\,1.957$\,\mu$m/$\lambda$\,2.121$\,\mu$m and H$_2\lambda$\,2.033$\,\mu$m/$\lambda$\,2.223$\,\mu$m
vs.\ H$_2\lambda$\,2.247$\,\mu$m/$\lambda$\,2.121$\,\mu$m    (Fig.\ \ref{fig2}) --- \citet{mouri94} proposed these diagrams
to separate gas emission yielded by shocks from emission caused by fluorescence. The drawback in using
the H$_2\lambda$\,1.957$\,\mu$m/$\lambda$\,2.121$\,\mu$m ratio is that the H$_2\lambda$\,1.957$\,\mu$m may be affected
by telluric bands of $\rm H_{2}O$ and $\rm CO_{2}$, or blended with
the [Si\,{\sc iv}]\,$\lambda$\,1.963$\,\mu$m emission line \citep{rodriguez-ardila05}. 

\item $ $ [O\,{\sc iii}]\,$\lambda\,5007$\AA/H$\beta$ vs.\ 
[O\,{\sc i}]\,$\lambda$\,6300\AA/H$\alpha$    (Fig.\ \ref{fig4}) --- This diagram was  
suggested by  \citet{baldwin81} to separate objects according to their primary excitation mechanisms, i.e.
(a) photoionization by stars, (b) photoionization by a power law continuum source or (c) shock heating.
In particular, the  [O\,{\sc i}]\,$\lambda$\,6300\,\AA/H$\alpha$ line ratio is
 greatly increased by the presence of shocking gas, even when it has low velocities (e.g. \citealt{allen08}).

\end{itemize}

\section{Results}
\label{res}

\subsection{Integrated spectra}

In Figs.\ \ref{fig1} and \ref{fig2}   we show the first three diagnostic
diagrams described above containing the results of our grid of
photoionization models and the data sample. Sy\,1, Sy\,2 and quasar are
represented by different symbols. For the IFU data,   the emission line
ratios represented in these Figs.\ were estimated by integrating the spaxels
inside a central aperture of 0.5\arcsec$\times$0.5\arcsec\ for each galaxy,
with exception of ESO428-G14 for which an aperture of
0.75\arcsec$\times$0.75\arcsec\ was considered.  These values are presented in Table~\ref{table2a}.

In Fig.~\ref{fig1}, we can see that almost all the observational ratios are
within the parameter space defined by our grid of photoionization models. A
lower metallicity than those assumed in our models is required to reproduce
the data of the galaxies  out of the
grid. The observed correlation between [Fe\,{\sc
ii}]\,$\lambda$\,1.2570$\,\mu$m/Pa$\beta$ and 
H$_2$\,$\lambda$\,2.1218$\,\mu$m/Br$\gamma$ is explained by an increase in
metallicity and ionization parameter. Noteworthy that the parameter space
defined by the models built using $Z\,=\,0.5\,Z_{\odot}$ is almost completely
contained in the one defined by the models built using the solar metallicity.

\begin{table}
\centering
\caption{Integrated line ratio intensities of IFU data}
\label{table2a}
\vspace{0.2cm}
\begin{tabular}{l c c }
\hline
\vspace{0.1cm}
     Object     &    [Fe\,{\sc ii}]\,$\lambda$\,1.2570$\,\mu$m/Pa$\beta$            &   H$_2\,\lambda$\,2.1218$\,\mu$m/Br$\gamma$  \\	
  ESO428-G14            &      0.75   &   1.10 \\
  Mrk\,1066                  &     0.52    &   0.96  \\
  Mrk1\,157                &    0.73  &   2.24  \\
  NGC\,4151             &    0.45   &   0.26  \\
\hline
\end{tabular}
\end{table}

In the case of the diagnostic diagrams that only involve H$_{2}$ line ratios
(Fig.~\ref{fig2}), the photoionization models are slightly dependent on the
assumed metallicities, covering almost the same parameter space, and 
strongly dependent on variations in the ionization parameter.   Taking into 
account the observational error bars, our models are in good agreement with the
observed  H$_2\lambda$\,2.033$\,\mu$m/$\lambda$\,2.223$\,\mu$m
and H$_2\lambda$\,2.247$\,\mu$m/$\lambda$\,2.121$\,\mu$m ratios (upper
panel). On the other hand, in the lower panel of  Fig.~\ref{fig2}  can
be noticed larger
dispersion of the observational data which is not well reproduce by the
models. This dispersion could be the result of a contamination of the
measurements of the H$_2\,\lambda$\,1.957\,$\mu$m emission line intensities
due to a blend with the [Si\,{\sc iv}]\,$\lambda$\,1.963$\,\mu$m line (as
explained above). Therefore, the predicted H$_2\lambda$1.957$\,\mu$m
intensities are somewhat lower than the observed ones.  
  In Fig.~\ref{fig2}, we also show the area occupied by the theoretical
intensities of the line ratio
H$_2\lambda$\,2.247$\,\mu$m/$\lambda$\,2.121$\,\mu$m from   shock models performed by 
\citet{hollenbach89}. These authors computed emission-line spectra of steady interstellar shocks
in molecular gas considering velocities from 30 to 150 km/s and particle
densities of $10^{3}-10^{6} \: \rm cm^{-3}$.  We can see that most of the
objects of our sample appear to have  the  X-rays as main ionizing source
while  for the  remaining ones a composite ionization
by  X-rays and shock  can be considered.
 
\begin{figure}
\centering
\includegraphics[angle=0,width=9.0cm]{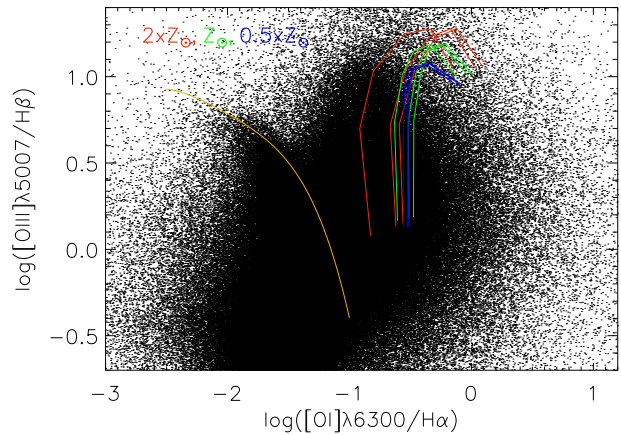}
\caption{[O\,{\sc iii}\,$\lambda$\,5007\,\AA/H$\beta$ vs.\ [O\,{\sc
   i}\,$\lambda$\,6300\,\AA/H$\alpha$ diagnostic diagram. The yellow line
   separates objects ionized by massive stars from   those
   containing active nucleus \citep{kewley01}.  Blue, green and red solid lines
   are as in Fig.\ \ref{fig1}. Points represent emission-line galaxies listed
   in the MPA/JHU Data catalogue of the Sloan Digital Sky Survey DR7 release
   (see Sect. \ref{obs}). }
 \label{fig4}
 \end{figure}

Fig.~\ref{fig4} shows the [O\,{\sc iii}]\,$\lambda$\,5007\,\AA/H$\beta$
vs.\ [O\,{\sc i}\,$\lambda$\,6300\,\AA/H$\alpha$  
diagnostic diagram. 
In this Fig.\ we can see that the observational data of AGNs are well describe
by our models.  If our models use the lower values of the ionization
parameter ($\log\,U\,<\,-3.5$; these models are not shown), we can extend the
parameter space to include the objects that  have values of the
logarithm of the [O\,{\sc iii}]\,$\lambda$\,5007\,\AA/H$\beta$ ratio lower
than zero. As in the case of the [Fe\,{\sc
ii}]\,$\lambda$\,1.2570$\,\mu$m/Pa$\beta$ and
H$_2$\,$\lambda$\,2.1218$\,\mu$m/Br$\gamma$ diagnostic diagram
(Fig.\ \ref{fig1}), the parameter space of the models with solar metallicity
almost contains that of the $Z\,=\,0.5\,Z_{\odot}$ models.

 \subsection{IFU data}

\begin{figure*}
\centering
\includegraphics[scale=0.9]{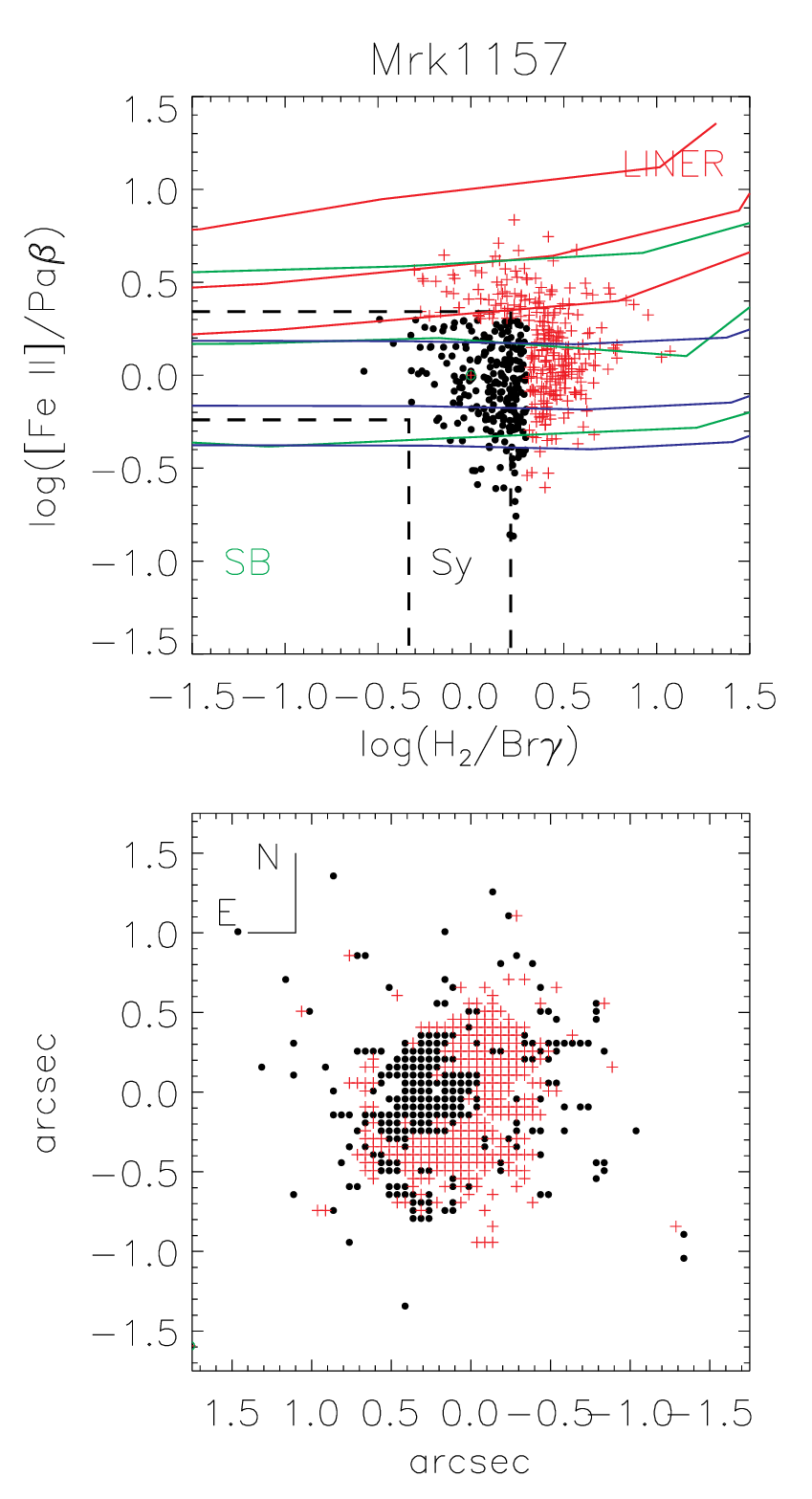}
\includegraphics[scale=0.9]{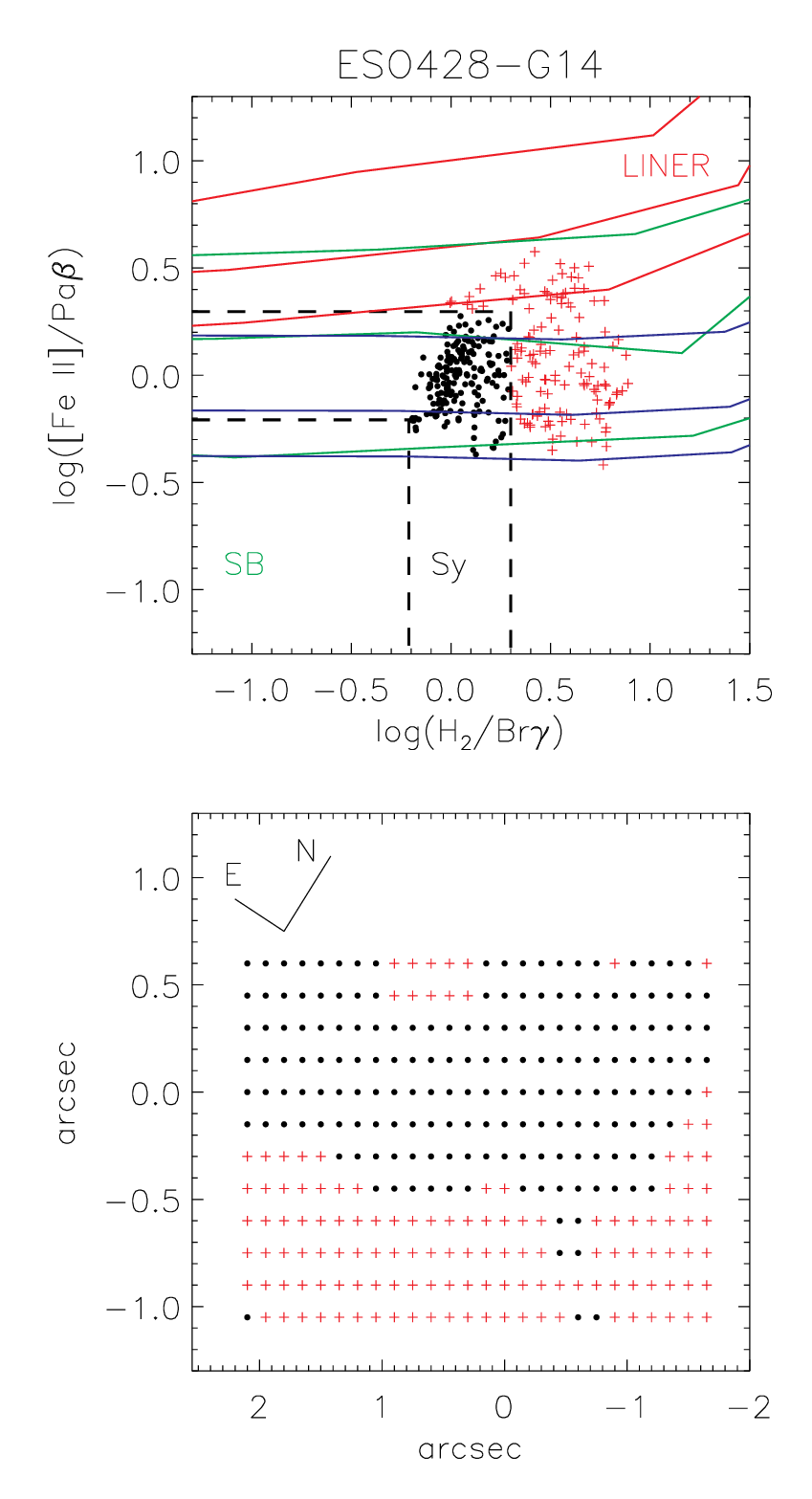}
\caption{Top panels: [Fe\,{\sc ii}]\,$\lambda\,1.2570\,\mu$m/Pa$\beta$ vs.\
H$_2$\,$\lambda\,2.1212\,\mu$m/Br$\gamma$ line-ratio
diagnostic diagram for   Mrk\,1157 (left) and ESO\,428-G14 (right). 
The dashed lines delimit regions with ratios typical of
Starbursts (green open circles), Seyferts (black filled circles) and LINERs (red
crosses).  Blue, green and red solid lines are as in Fig.\ \ref{fig1}. Bottom
panels: spatial position of each spaxel in the IFU field of view from the
diagnostic diagram.} 
\label{fit1} 
\end{figure*}

\begin{figure*}
\centering
\includegraphics[scale=0.9]{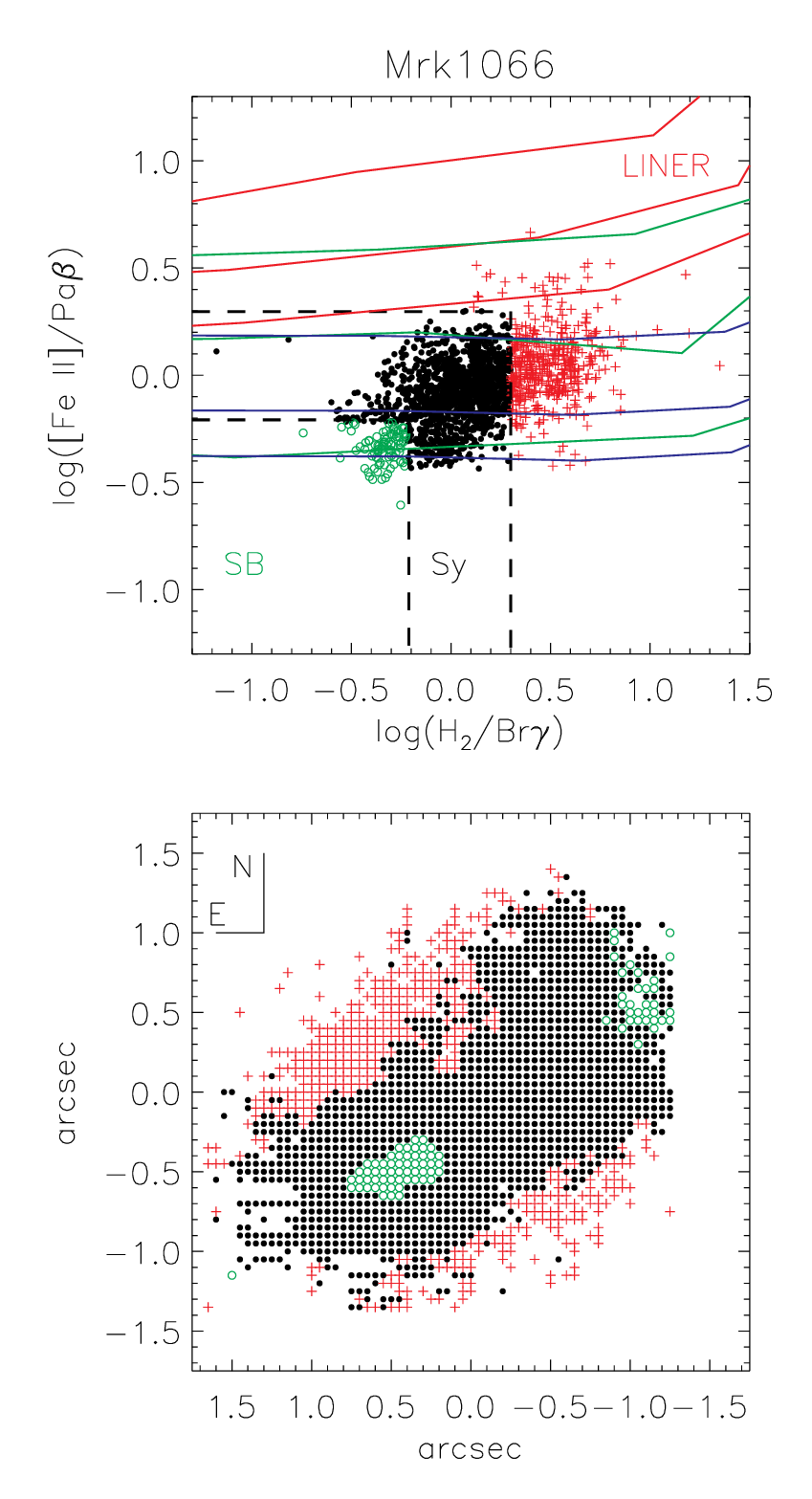}
\includegraphics[scale=0.9]{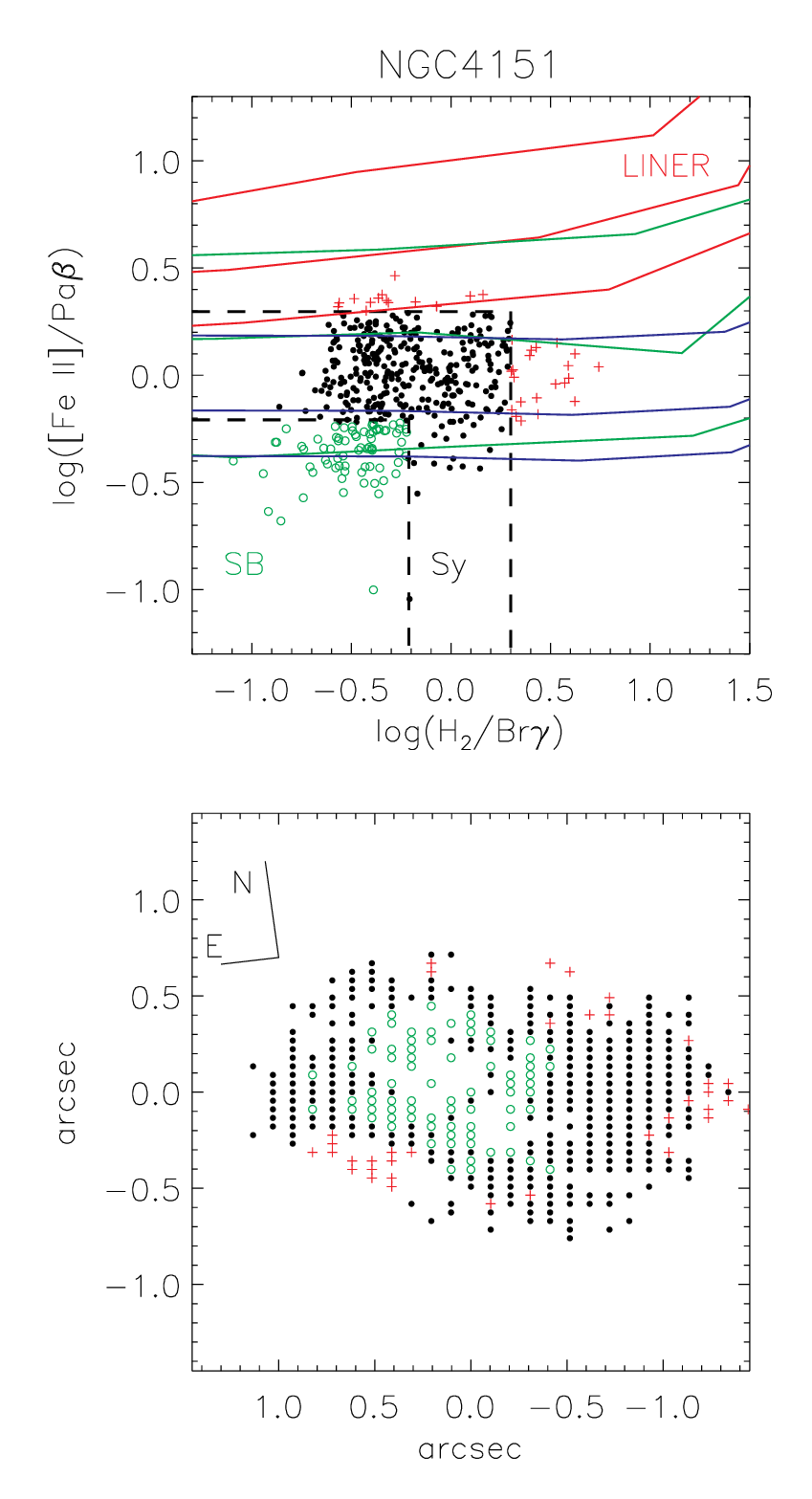}
\caption{As in Fig.\ \ref{fit1} but for Mrk\,1066 (left) and NGC\,4151
    (right).}
\label{fit2} 
\end{figure*}

We plot the [Fe\,{\sc ii}]\,$\lambda$\,1.2570$\,\mu$m/Pa$\beta$ and
H$_2$\,$\lambda$\,2.1218$\,\mu$m/Br$\gamma$ diagnostic diagram for each spaxel 
of our four objects with our model results (see upper panels of
Figs.\ \ref{fit1} and \ref{fit2}). In these Figs.,  the spaxel data are separated by their
ionization mechanism according to the place in the diagnostic diagram,
 with different colours for each mechanism.  
The different ionization mechanism zones are delimited in the Figs.  by
dashed-lines, following the work of \cite{rodriguez-ardila04}. The spaxels
showing typical values of starbursts, Seyferts, and low-ionization nuclear
emission-line regions (LINERs) are represented by green open circles, black
filled circles, and red crosses, respectively.  With the same colour code,
we show the spatial position of each spaxel in the IFU field of view (see
lower panels of Figs.\ \ref{fit1} and \ref{fit2}). 
Our models completely represent the region occupied by Seyfert and LINERs data.

\section{Discussion}
\label{disc}

The excitation mechanism of the near-IR emission lines of the [Fe\,{\sc ii}]
and  H$_2$   in active galaxies have been the subject of several works.  
 For example, \cite{mouri00} compared results of models considering
photoionization by X-rays and shock heating with  observational data of
AGNs and starburst galaxies. These authors built their models considering
large ranges in shock velocities, gas density, metallicity, and different
ionizing continua. Mouri and collaborators showed that the [Fe\,{\sc ii}]
emission is enhanced when a partially ionized zone is produced by
photoionization by X-rays (described by a power-law) and shock heating. These
two processes can be discriminated by the electron temperature of the 
[Fe\,{\sc ii}] region: 8000\,K in heating by X-rays and 6000 K in shock
heating. Comparing the electron temperature of the [Fe\,{\sc ii}] region
estimated by \citet{1995ApJ...454..660T} for NGC\,4151 ($8000\: < \: T_{\rm e}
\: < \: 12000$\,K) with their models, \cite{mouri00} showed that, at least for
this galaxy, it indicates that X-rays are the more important mechanism to
yield the [Fe\,{\sc ii}] flux. These authors arrived to the same conclusion
using the [O\,{\sc i}]\,$\lambda$\,6300\,\AA/H$\alpha$ vs.\ [Fe\,{\sc
ii]$\lambda$1.2570$\,\mu$m]/Pa$\beta$ diagnostic diagram. 
A similar result was also obtained by \cite{jackson07}  by  analysing J-band
spectra of  three Sy\,2 galaxies.} 

\begin{figure}
 \centering
 \includegraphics[width=.95\columnwidth, angle=0]{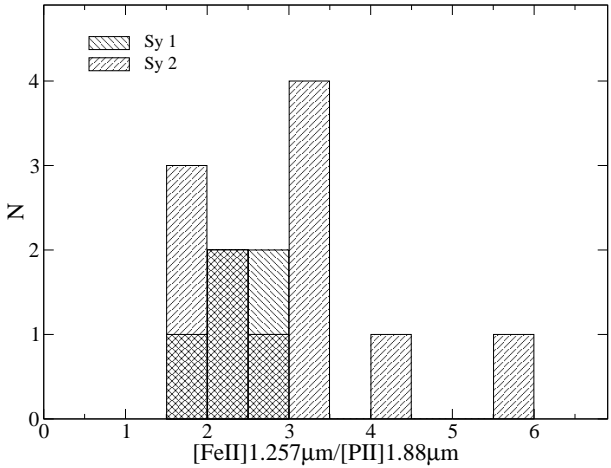}
 \caption{   Histogram showing the [Fe\,{\sc ii}]/[P\,{\sc ii}]
  emission line intensity ratios of a sample of objects collected from the
  literature.} 
 \label{phosp}
 \end{figure}

For our models, we assumed an incident continuum whose shape is given
by two components, a Big Bump and an X-ray power law, varying the Fe/O
abundance. With these models that do not consider shock heating, we are able
to explain the observational data. Nevertheless, we do not exclude some
contribution by shock heating to the [Fe\,{\sc ii}] emission. Comparing our
models with  the SDSS DR7 emission-line galaxies (Fig.\ \ref{fig4}), 
we are able to describe the [O\,{\sc
i}]/H$\alpha$ line ratio observed in AGNs. This diagram cannot be used to
discriminate [Fe\,{\sc ii}] excitation mechanism, nevertheless, we must take
into account that the [O\,{\sc i}]/H$\alpha$ line ratio is shock sensitive. Hence, 
although shock contribution in the ionization of Fe cannot be ruled out,  
models considering a continuum described by a Big Bump and an X-ray power law
as the ionization source 
can also reproduce the [Fe\,{\sc ii}] emission lines as well as the behaviour of
shock sensitive emission lines such as [O\,{\sc i}]\,$\lambda$\,6300\,\AA.
Analysing IFU observations of the Sy galaxy NGC\,4151, \citet{turner02} found
that the [Fe\,{\sc ii}] emission mainly arises in the visible narrow-line
region in which the dominant excitation mechanism is the photoionization by collimated X-ray
emission from the nucleus.
  \citet{oliva01} pointed out that in regions where shocks are the dominant mechanism
the iron-based grains are destroyed but the phosphorus is not,   
yielding  a larger  [Fe\,{\sc ii}]\,$\lambda\,1.2570$\,$\mu$m/[P\,{\sc ii}]\,$\lambda\,1.188$\,$\mu$m
line-ratio  intensity than that observed in  the regions 
dominated by X-ray.
  In order to
verify this, in Fig.~\ref{phosp} we show a histogram containing this observed
line intensity ratio for 17 Seyfert galaxies, 5 Sy\,1 and 12 Sy\,2, taken from
\citet{jackson07},  \citet{rogerio06}, and \citet{oliva01}. It can be seen that the [Fe\,{\sc ii}]/[P\,{\sc ii}] for Sy galaxies ranges from 1.5 to 6 (with a
mean value of 2.7). The mean value of this ratio is about  20 for SNRs, which
indicate that the emitting gas has recently passed through a fast shock
\citep{oliva01}. Therefore,  these results confirm that shocks have little
influence on the [Fe\,{\sc ii}] emission.

Regarding the $\rm H_{2}$, this molecule can be excited via three distinct mechanisms:
(1) UV fluorescence, where photons with $\lambda\,>\,912$\,\AA\
are absorbed by the  $\rm H_{2}$ molecule and then re-emitted,
resulting in the population of various  vibro-rotational levels, (2) shocks,
where high-velocity gas motions 
heat and accelerate   this molecule; and (3) X-ray illumination, where hard X-ray
photons penetrate deep into molecular clouds, heating large
amounts of molecular gas resulting in the $\rm H_{2}$ emission \citep[see][and
references therein]{rodriguez-ardila04}.  Rodr\'iguez-Ardila and
collaborators used the diagrams shown in Fig.~\ref{fig2} to compare
observational data of 22 objects with models considering a thermal emission, a
non-thermal UV excitation, a thermal UV excitation, and a mixture of thermal
and low-density fluorescence. 
These authors found that for 4 objects the excitation mechanism is clearly
thermal, while for the remaining  objects a mixing with a non-thermal
process cannot be discarded,  even though the results point out to a
dominant thermal mechanism.

\begin{figure}
 \centering
 \includegraphics[height=.95\columnwidth, angle=-90]{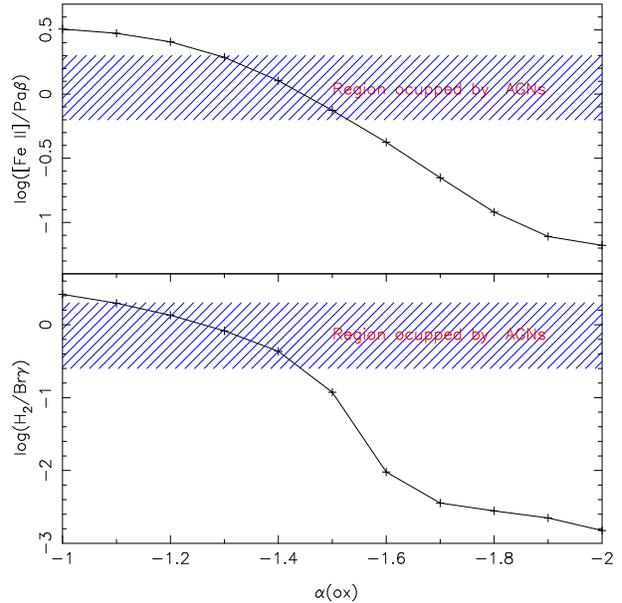}
 \caption{Model results using solar metallicity, N$_e$\,=\,$10^4$\,cm$^{-3}$,
     log\,U\,=\,-2.5 and varying only the $\alpha_{ox}$ parameter to 
     see the influence of the X-rays on the 
     [Fe\,{\sc ii}]\,$\lambda\,1.2570\,\mu$m/Pa$\beta$ and
     H$_2$\,$\lambda\,2.1218\,\mu$m/Br$\gamma$ emission line ratios. To
     delimit the region occupied by the AGNs we follow \citet{rodriguez-ardila04}.}
 \label{vari-alphaox}
 \end{figure}

  To analyse the relative weight of the X-ray emission with respect to the
other model components (mainly with fluorescence and UV photons), not only for
the  $\rm H_{2}$ emission but also for 
the [Fe\,{\sc ii}], we made models fixing all parameters with the exception of
the $\alpha_{ox}$ value (see Fig.\ \ref{vari-alphaox}), which is related to
the X-ray power law normalization (see Section \S \ref{phot}). We assumed
$Z$\,=\,$Z$$_\odot$ and log\,$U$\,=\,$-$2.5 since the models built using the solar
metallicity and this value of the ionization parameter cover
almost all the parameter space occupied by the observational data (see
Fig.\ \ref{fig1}).
Taking into account the $\alpha_{ox}$ definition \citep{tananbaum79}, which fixes
the Big Bump parameters, a decrement of the $\alpha_{ox}$ value implies that
the amount of the X-rays emitted by the source decreases. 
In Fig.\ \ref{vari-alphaox} we can see that our models with $\alpha_{ox}=-1.4$
reproduce well the observational AGN data. Nevertheless, when we  use lower values
of this parameter, the ratios predicted by the models go out of the region
typically occupied by the AGNs \citep{rodriguez-ardila04}. Therefore, our
models favour the scenario 
suggested by \citet{maloney96}, where the $\rm H_{2}$ molecule    emission
is mainly governed by photons emitted at X-ray wavelengths from the central AGN.
This also can be inferred from the dependence of the $\rm H_{2}$ emission lines 
on the ionization parameter $U$. 

  To verify if shock models can fit the observational data, we compared shock
model results by \citet{hollenbach89} with our sample (see
Fig.~\ref{fig2}). Only few observational points are located in the area
occupied by these shock models and, even in these cases, models considering
X-rays also describe the data. 

  On the other hand, varying in our models the $\rm H_{2}$ ionization
rate by cosmic rays by a factor of 200, we found that the
H$_2\,\lambda$\,2.1218$\,\mu$m/Br$\gamma$ line ratio only increases by about
0.15\,dex, which shows that the additional ionization by cosmic rays has
little influence on the $\rm H_{2}$ emission lines.

\begin{figure}
 \centering
 \includegraphics[width=.95\columnwidth]{SED-mkn1066-data.eps}
 \caption{Spectral energy distribution at the Schwarzschild radio
   (10$^{-5}$\,pc) of the Sy\,1 galaxy Mrk\,1066 used as the photoionization
   source for some models of this galaxy. We assumed a galaxy distance of
   50\,Mpc \citep{mould00}. The photometric data were taken from
     \citet{1978ApJS...36...53D},  
     \citet{1990BAAS...22Q1325M},
     \citet{1991trcb.book.....D},
     \citet{1991ApJS...75....1B},
     \citet{1996AJ....111.1945D},
     \citet{2002AJ....124..675C},  
     \citet{2003yCat.7233....0S},  
     \citet{2004ApJ...617L..29B}, 
     \citet{2005A+A...444..119G},
     \citet{2007AJ....134..648M},
     and
     \citet{2007AJ....134.1263C}.
     } 
 \label{fig5}
 \end{figure}

 \begin{figure}
 \centering
 \includegraphics[angle=-90,width=8cm]{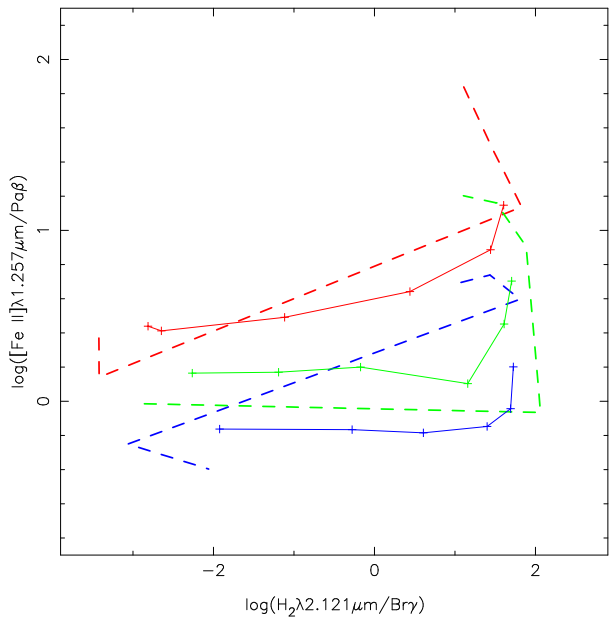}
 \caption{Comparison between the grid model results shown in Fig.\ \ref{fig1}
   (solid lines) and the models built considering 
 the semi-empirical SED of Mrk\,1066 shown in Fig.~\ref{fig5} (dashed lines).}
 \label{fig6}
 \end{figure}
  
   A simple scenario where both [Fe\,{\sc ii}] and $\rm H_{2}$  emissions are
mainly due to the X-ray continuum coming
from the active nucleus has also been proposed by other authors. 
For example, \cite{blietz94} and \citet{knop01} showed that
X-rays from the nucleus can heat the gas located in the narrow line region
driving the [Fe\,{\sc ii}] and H\,$_2$ emission.  Because 98\,\% of the iron is tied up in
dust grains, this process must free the iron through dust destruction
and yet not destroy the $\rm H_{2}$ molecules \citep{rodriguez-ardila04}.    
These authors computed the 
emergent [Fe\,{\sc ii}]\,$\lambda$\,1.2570\,$\,\mu$m and  H$_2\,\lambda$\,2.1218$\,\mu$m
flux using the X-ray models by \citet{maloney96} and compared their predictions
with observational data of seven objects. They found that X-ray heating
can only explain a fraction of the [Fe\,{\sc ii}] and H$_2$ emission, and they
  stated that the discrepancy found can be alleviated if the emitting gas
is located closer than the distance adopted   in their models.
  The X-ray data, provided by the \textit{XMM-Newton} and \textit{Chandra}  
space telescopes, and their detailed analysis
  \citep[see e.g.\ ][and references therein]{piconcelli05,longinotti07,bianchi09,krongold09,cardaci11,corral11},
 provide information about the continuum shape and the particular
spectral features of the AGNs in this wavelength range.
For Mrk\,1066, we compared the results obtained using this simple scenario
that only involves a continuum modelled by a Big Bump and an X-ray power law
with those obtained using its intrinsic SED. We built the observational SED
taking the photometric data from the NASA/IPAC Extragalactic Database (NED),
following \citet{cardaci09}.  
To enhance the number of points of the SED as needed by Cloudy, we
performed a linear interpolation among the semi-empirical points (see
Fig.\ \ref{fig5}).
We built a new grid of photoionization models under the same assumptions of
abundances, ionization parameters and density, but only for one value of the
Fe/O ratio for each metallicity. 
In Fig.~\ref{fig6} the predictions of our models using the SED of Mrk\,1066
and the model results presented in Fig.~\ref{fig1}  
assuming the same Fe/O abundance  as the Mrk\,1066 models
are shown. The model results   derived using the two
different ionizing sources are mostly in agreement.

The semi-empirical SED of Mrk\,1066 includes not only the range covered by the
Cloudy model but also the radio and IR wavelengths. 
Hence, the agreement between solid and dashed lines in Fig.~\ref{fig6} only
indicates that the assumed multicomponent model is a good representation of
the AGN continuum when studying the [Fe\,{\sc ii}] and H$_2$ emission. 

Recent resolved integral field spectroscopy of the central region of
active galaxies shows that the ionized (in particular the [Fe\,{\sc ii}]
emitting gas) and the molecular (traced by the H$_2$ emission) gas have
distinct flux distributions and kinematics.   The molecular component is more
restricted to the plane of the galaxies and the ionized one extends to high
latitudes above it,  which is in  most cases co-spatially with the radio jet
\citep[e.g.,][]{eso428,riffel08,riffel09,m1066,riffel11a,riffel11b,n4151,thaisa10}. 
Usually the  [Fe\,{\sc ii}] has an enhancement in flux and velocity
dispersion in regions  surrounding   the radio structure,
suggesting that the radio jet plays an important role in the [Fe\,{\sc ii}]
emission. Our models are able to reproduce the [Fe\,{\sc ii}] emission of
active galaxies without considering shock excitation by the radio jet. Thus,
the enhancement in the [Fe\,{\sc ii}] flux in the vicinity of radio
structures can be interpreted as being due to an enhancement in the gas
density, caused by the interaction of the radio jet with the emitting gas, and
mainly excited by X-rays from the central engine.
 
   The main exciting mechanism of infrared emission lines of
ESO\,428-G14, Mrk\,1157, Mrk\,1066 and NGC\,4151 was discussed by
\citet{eso428,riffel08,m1066} and \citet{n4151}, respectively. However, these
authors did not reach conclusive results. For example, \citet{eso428}
suggested that the [Fe\,{\sc ii}] excitation in ESO\,428-G14 is 
mainly due to shocks. Nevertheless, the detailed analysis performed in the
present work confronting our models with the IFU data shows that X-rays are a
more reliable dominant excitation mechanism even in the case of ESO\,428-G14.

\section{Conclusions}
\label{conc}
 
In this work we show that a photoionization model grid built by adopting a
continuum source characterized by two components, one accounting for the
Big Bump component peaking at $\rm 1 \: Ryd$ and the other describing the
X-rays emission, is able to reproduce the [Fe\,{\sc ii}] and  H$_2$ infrared
emission lines of a sample of AGNs. 
Testing the influence of the X-rays on the intensity of these emission lines,
we found that a decrement in the X-ray content of the continuum source
translates into a weakening of these lines, and the models are no longer 
compatible with the observations.
This implies that the heating by   the X-ray emission from the  active nuclei
can be considered as the   most important mechanism of excitation for the IR
emission lines of these elements.

\section*{Acknowledgments}
  We are grateful to the referee, Neal Jackson, for a thorough reading of the
manuscript and for suggestions that greatly improved its clarity.

Based on observations obtained at the Gemini Observatory, 
which is operated by the Association of Universities for Research in Astronomy, Inc., under a cooperative agreement with the 
NSF on behalf of the Gemini partnership: the National Science Foundation (United States), the Science and Technology 
Facilities Council (United Kingdom), the National Research Council (Canada), CONICYT (Chile), the Australian Research 
Council (Australia), Minist\'erio da Ci\^encia e Tecnologia (Brazil) and south-eastCYT (Argentina).  
This research has made use of the NASA/IPAC Extragalactic Database (NED) which
is operated by the Jet Propulsion Laboratory, California Institute of  Technology, under contract with the National Aeronautics and Space Administration.
OLD and ACK are grateful to the FAPESP for support under grant 	
2009/14787-7 and 2010/01490-3. MC and GH are grateful to the Spanish Ministerio de Ciencia e Innovaci\'on for
support under grant AYA2010-21887-C04-03, and the Comunidad de Madrid under 
grant S2009/ESP-1496 (ASTROMADRID). EPM is grateful to
the Spanish Ministerio de Ciencia e Innovaci\'on for support under
grant AYA2010-21887-C04-02, and the Junta de Andaluc\'ia under grant TIC114.

\bibliographystyle{mn2e}
\bibliography{references}

\label{lastpage}

\end{document}